\documentclass[twocolumn]{article}
\usepackage{usenix,epsfig,url}
\begin{document}

\date{}

\title{\Large \bf On The Cost Distribution of a Memory Bound Function}

\author{
David S.\ H.\ Rosenthal \\
{\em Stanford University Libraries} \\
{\em Stanford, CA 94305}\\
{\normalsize \url{http://www.lockss.org}} \\
}

\maketitle

\thispagestyle{empty}

\subsection*{Abstract}

Memory Bound Functions have been proposed for fighting spam,
resisting Sybil attacks and other purposes.
A particular implementation of such functions has been proposed
in which the average effort required to generate a proof of effort
is set by parameters $E$ and $l$ to $E \cdot l$.
The distribution of effort required to generate an individual
proof about this average is fairly broad.
When particular uses of these functions are envisaged,
the choice of $E$ and $l$, and the system design surrounding
the generation and verification of proofs of effort,
need to take the breadth of the distribution into account.

We show the distribution for this implementation,
discuss the system design issues in the context of two
proposed applications,
and suggest an improved implementation.

\section{\label{sec:introduction}Introduction}

Following Abadi et al\cite{Abadi2003},
Dwork et al~\cite{Dwork2003} suggest imposing computational costs
on the senders of e-mail by requiring the mail to be accompanied
by a \emph{proof of effort} generated using a class of Memory Bound
Functions (MBF).  
Specifically,
they propose a concrete implementation called \emph{Algorithm Mbound}.
The average computational effort, measured in
cache misses, it requires to generate a proof of this kind is set by
two parameters $E$ and $l$.  Generating a proof requires on average
$E \cdot l$ cache misses,  verifying it requires $l$ cache misses.

The distribution of effort required to generate an individual
proof about the average of $E \cdot l$ is fairly broad.
When particular uses of MBF are envisaged,
the choice of $e$ and $l$, and the system design surrounding
the generation and verification of proofs of effort,
need to take the breadth of the distribution into account.

We show this distribution and examine the resulting design issues
in the context of the use of MBF both to impede spammers,
and to assist the
LOCKSS\footnote{LOCKSS is a trademark of Stanford University.}
(Lots Of Copies Keep Stuff Safe) system in resisting
attacks~\cite{Maniatis2003lockssSOSP}.
We then propose an MBF implementation that has a better
generation cost distribution and other advantages.

\section{\label{sec:mbfScheme}Memory Bound Functions}

In this section we introduce Memory Bound Functions,
and describe their application in spam prevention
and in the LOCKSS protocol.

\subsection{\label{sec:mbfScheme:Background}Background}

The goal of a Memory Bound Function is to cause the generator of
a proof to incur $C$ cache misses and thus RAM accesses.  If
each of these takes $t$ seconds, the generator must have used $C \cdot t$
seconds on a real computer.

Algorithm MBound has two adjustable parameters, the cost of verifying
an effort proof $l$ and the ratio, $E$, between $l$ and the cost of
constructing the proof.  We measure all costs in cache misses, so a
proof costs $E \cdot l$ cache misses to construct and $l$ cache misses
to verify.  It uses an incompressible fixed public data set $T$ larger
than any cache it is likely to meet.  A proof generator who must
expend effort $E \cdot l$ is given as challenge a nonce $n$ (so that
older effort proofs cannot be reused) and the values of $E$ and
$l$.  In response, the generator must perform a series of pseudo-random
walks in the table $T$.  Each starting point is determined by an index $s$
which starts at zero and is incremented for each walk, or trial.
The walk computes a one-way value
$A$ based on $n$, $s$ and the encountered elements of table $T$.
The walk is dependent on $n$ and $s$;
it is constructed so that the number of encountered elements is
$l$, and fetching each encountered element causes an L1 cache
miss. Each walk, therefore, causes $l$ cache misses.

The generator stops when the value $A$ computed by the walk has 0 bits in its
least significant $e = \log_2E$ positions.  It is thus
expected the generator will try on average $2^e$ walks with different starting
positions determined by successive values of $s$ before finding an appropriate
starting position $i$,
costing $C = E \cdot l$ cache misses.

The $i$ that yielded the appropriate $A$ is the effort proof.  The
verifier need only perform the random walk on $T$ starting with the $n$
he chose and the $i$ sent by the generator, costing $V = l$ cache misses.
If the resulting $A$ has the proper 0-bits in its last $e$
positions, the verifier accepts the effort.

There is some non-zero probability that the proof search will take longer
than \emph{any} given number of steps.
Dwork et al acknowledge this indirectly when they say:
\begin{quote}
``If $i > 2^{2e}$ the
[verifier] rejects the message (with overwhelming probability one of the
first $2^{2e}$ trials should be successful).''~\cite{Dwork2003}
\end{quote}

Indeed,
but what of the poor proof generator who had,
using their parameters,
expended more than $2^{42}$ cache misses (about 10 days)
and still failed to find a proof?
It isn't just the verifier who must reject extremely long proofs,
but also the generator.
In any practical implementation of Algorithm MBound,
account needs to be taken of the fact that some proportion of the
attempts to generate a proof will fail simply because they took too long.

\subsection{\label{sec:mbfScheme:Spam}Use of MBF in Spam Prevention}

The idea of using MBF to prevent spam is for a receiver to refuse to
accept mail that is not accompanied by ``postage'' in the form of a
proof of effort based on the sender's alias, the receiver's alias
and the body of the message. The average amount of effort would be set to
prevent an individual machine, even the fastest available, sending
more than a few thousand mails per day.  The goal of the spam adversary
would be to reduce the actual effort required to send spam below this
average.

\subsection{\label{sec:mbfScheme:LOCKSS}Use of MBF in LOCKSS}

The LOCKSS system implements a peer-to-peer network of persistent,
muutally suspicious web caches used to preserve access to e-journals.
The LOCKSS protocol requires a peer A requesting a service from a peer B
to obtain a nonce from B,
then provide B with a proof of effort based on that nonce representing more
effort than B will need to perform the requested service.
B verifies this effort before performing the service and returning the
result to A.  In the LOCKSS protocol, the service peers provide each other
is that of voting on the digest of blocks of content.  The proofs of
effort prevent a conspiracy of a minority of malign peers overwhelming
these votes, or cheaply swamping the system in bogus polls.

\section{\label{sec:costDistribution}Distribution of Proof Generation Effort}

\begin{figure}
\centerline{\includegraphics{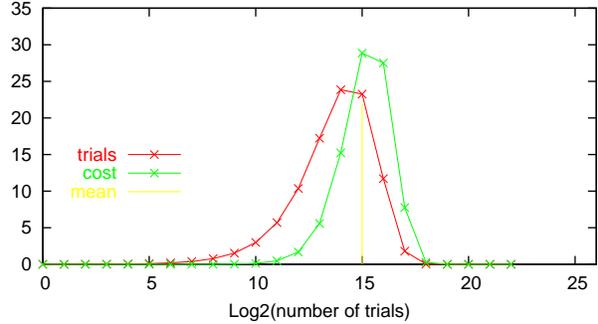}}
\caption{Histograms of number of trials and percentage of total cost
for $e = 15$}
\label{fig:e15Histogram}
\end{figure}

Figure~\ref{fig:e15Histogram} shows the behavior of Algorithm MBound using
the authors' choice of $e = 15$.
We show histograms of the percentage of proof generations whose search ended
after a given number of trials,
and of the percentage of the total cost of proof generation due to those
proofs.
Note the skewed distributions;
the majority of proofs take less than expected but,
because they are cheap,
the majority of the cost is due to proofs taking longer than expected.

\begin{figure}
\centerline{\includegraphics{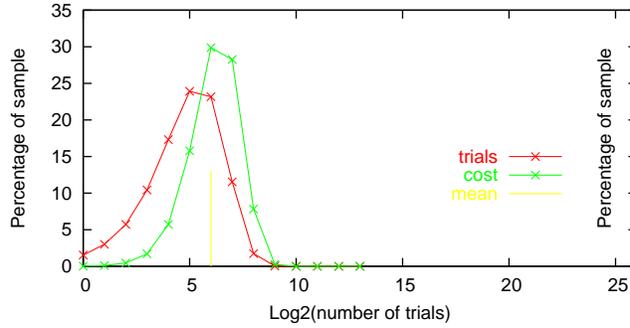}}
\caption{Histograms of number of trials and percentage of total cost
for $e = 6$}
\label{fig:e6Histogram}
\end{figure}

From this data we see that:
\begin{itemize}
\item Even though the average cost of a proof is 32768 trials,
more than 63\% of the proofs are found in less than 16384 trials.
The median proof lies between 8192 and 16384 trials,
\item There is almost 0.1\% probability of finding a proof in 32 trials or less.
\item There is almost 12\% probability of finding a proof in 4096 trials or less.
\item 9\% of the total cost represents proofs taking more than four times
as long as expected.
\item 40\% of the total cost represents paths taking more than twice as long
as expected.
\end{itemize}

\begin{figure}
\centerline{\includegraphics{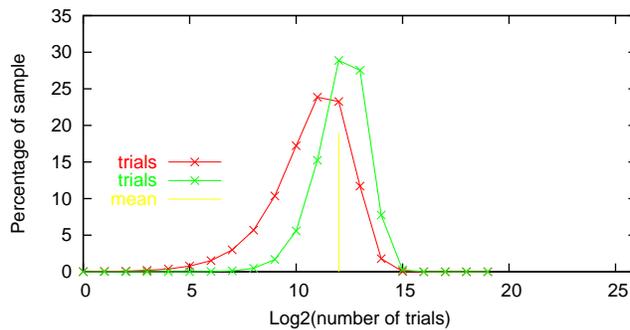}}
\caption{Histograms of number of trials and percentage of total cost
for $e = 12$}
\label{fig:e12Histogram}
\end{figure}

\begin{figure}
\centerline{\includegraphics{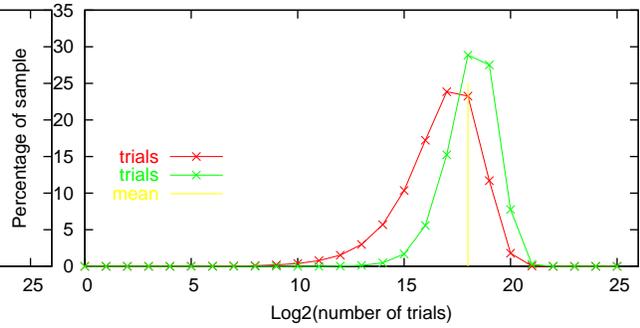}}
\caption{Histograms of number of trials and percentage of total cost
for $e = 18$}
\label{fig:e18Histogram}
\end{figure}

This behavior is not an artifact of the particular value of $e$ they choose.
Figures~\ref{fig:e6Histogram}, \ref{fig:e12Histogram}, \ref{fig:e18Histogram}
show the behavior for $e = 6, 12, 18$ respectively.

\section{\label{sec:designIssues}Design Issues}

In this section we discuss the issue raised by this distribution of
effort proof costs for both the spam prevention and LOCKSS protocol
applications.

\subsection{\label{sec:designIssues:proof}What Does The Proof Prove?}

At first sight it might appear that after the fact,
the verifier of a proof knows how much effort was used to find it.
The proof $i$ in Algorithm MBound is,
after all,
a count of the number of trials.
Although the verifier finds that index $i$ produces a path terminating
with a value with $e$ low-order zero bits,
that does not prove that no path $j < i$ produces such a path.
The verifier does not know that the generator started with index 0.

It is important to observe that the proof shows only that the generator did
something that,
averaged over a large number of repetitions by a generator
using \emph{exactly} the procedure specified,
required approximately $2^e \cdot l$ path steps.

The actual number of steps for the proof in question is unknown;
it is quite likely to be less than $2^{(e-3)} \cdot l$.
The average number of steps can be strongly influenced by the point
at which the generator abandons the search as taking too long,
a parameter that the generator doesn't reveal to the verifier.
This uncertainty about the relationship between the requested and
actual effort is a problem for both cases.

\subsection{\label{sec:designIssues:tails}Tails of the Distribution}

The left tail of Figure~\ref{fig:e15Histogram} is a region in which the cost
of proofs curve is below the number of proofs curve.  If a generator
could manipulate things to increase the probability of being in this region
it could generate proofs more cheaply than the verifier expects.
Similarly,  the right tail is a region in which the cost of proofs curve
is above the number of proofs curve.  If a generator could manipulate
things to decrease the probability of being in this region
it could generate proofs more cheaply than the verifier expects.

Both tails of the cost distribution pose problems for the LOCKSS protocol:
\begin{itemize}
\item The significant population of low-cost proofs provides an adversary
with a good chance of finding a proof much faster than expected.
\item The significant proportion of the cost represented by the small
proportion of very long proofs allows an adversary to roughly halve his cost
with little impact on his strategy simply by terminating the search for a
proof early and failing to vote.
\end{itemize}

The LOCKSS protocol uses random challenges which appear to prevent 
attacks analogous to ``chosen plaintext'' attacks on encryption systems.
Unfortunately,
the fault-tolerance of the protocol allows the adversary to selectively
refuse to supply expensive proofs
and thus bias the ``plaintext'' towards cheap proofs.

Analogous difficulties arise in the application of Algorithm MBound
to spam prevention.
At first sight this case is also immune from a ``chosen plaintext'' attack,
since the proof must be based on the text of the e-mail,
its source and destination address,
and so on.
Unfortunately:
\begin{itemize}
\item Spammers already perturb the text of their messages with random
garbage their victims will ignore but which helps evade signature-based spam
filters.  They could choose a e-mail text and try many possible starting
indices in parallel.
If none produced a cheap proof,
they could perturb the e-mail text and try again.
\item Similarly,
no individual recipient of spam is important.
By terminating the search for a proof early;
the spammer loses a proportion of the intended recipients but reduces
his cost much more.
\end{itemize}

\subsection{\label{sec:designIssues:parameters}Adjusting Costs}

The LOCKSS protocol needs to vary the additional costs imposed via
the MBF scheme over a wide range to match the economic requirements
of the protocol to the cost of the underlying hashing.

The LOCKSS application requires considerable freedom in
setting costs,
whereas for best results 
Algorithm MBound requires the values of $e$ and $l$ not to be too small,
and incrementing $e$ increases the cost by a factor of two.
With these restrictions,
the choice of $e$ and $l$ can be a problem.

Even in the spam application,
setting prices of 0 (from known senders) or 1 (from unknown senders)
for all e-mail ``postage'' lacks flexibility.
An ideal design would allow the recipient to accept less ``postage'' from
senders it trusted more.

\subsection{\label{sec:designIssues:timingUncertainty}Timing Uncertainty}

To limit the possibility an adversary could ``time-shift'' and use
a single computer to generate many votes in a single poll,
the LOCKSS protocol sets deadlines by which a generator must deliver
a proof.
To do so,
it must adjust for the difference between fast and slow memory systems.
Dwork et al report this is a factor of 5.
In addition,
when using Algorithm MBound it must also adjust for the
possibility that the generator may find a proof in many fewer or
many more steps than expected.
This ``slop'' in the timing defeats the purpose of the deadlines.

This uncertainty is not a problem in the spam application;
since the recipient only sees the fixed time taken to verify the proof.

\section{\label{sec:requirements}Requirements for MBF}

The LOCKSS  protocol's requirements for a cost function were
stated~\cite{Maniatis2003lockssSOSP} as:

\begin{quote}
``First, it must have an adjustable cost, since different
amounts of additional effort are needed at different protocol steps.
Second, it must produce effort measurable in the same units as
the cost it adjusts (hashing in our case).  Third, the cost of
generating the effort must be greater than the cost of verifying it.''
\end{quote}

To these we can now add:
\begin{itemize}
\item The distribution of actual effort involved in generating proofs of
effort about the predicted mean cost must be narrow.
\item It must not be possible for an effort generator to significantly
reduce the average cost of proof generation by failing long proofs
selectively.
\item The proof must be concise,
not requiring much data to be transmitted.
\item The effort must be continuously adjustable.
\end{itemize}

In fact, we would like our proof to prove that the generator performed
\emph{exactly} $E \cdot l$ steps.

\section{\label{sec:proposedSolution}An Alternative MBF}

Consider a mechanism in which the generator is asked to produce not
a single index $i$ which results in a value with $e$ low-order zeros,
but a list of all $i$ in a specified range which result in
values with $e - m$ low-order zeros,  where $2^m$ is the expected
number of such values.  The verifier can cheaply verify some
indices randomly chosen from the list, if any are invalid the proof
is invalid.  The more indices are verified,  the more sure the
verifier is of the effort expended by the generator.  If the
verifier considers extra verification effort justified,
a search of some randomly chosen intervals between the indices
in the list that revealed indices resulting in values with the
required number of zeros would also render the proof invalid.
The generator doesn't know which indices and ranges the verifier
will check,  so cannot include bogus entries without risking
discovery.


In this way we force the generator to explore an entire range of
indices,  rather than stopping at the first one.  The number of
trials needed to generate a proof is fixed, because every
index in the requested range must be checked if the generator is
unwilling to risk the cheat being discovered.  This satisfies the first two of
our additional requirements.

There will be considerable variation in the cost of verifying
a proof,  depending on how sure the verifier wants to be that
the generator actually exerted the requested effort.

This mechanism has some attractive features:

\begin{itemize}
\item There is no uncertainty as to the amount of effort exerted
by the generator;
unless it is willing to risk detection it must perform exactly
the effort requested.
\item The verifier never performs more effort than the generator;
it can stop as soon as any trial is invalid.
\item The more effort the verifier performs,
the better the verification.
The amount of this effort can be adjusted
to match the verifier's suspicion of
the generator.
This property is valuable in both applications we consider:
\begin{itemize}
\item Spammers wish to conceal their true identity behind a screen of
proxies,
making it unlikely that a recipient would engage in repeated
transations with the same apparent spammer identity.
But a recipient normally has repeated transactions with many
legitimate e-mail senders;
reducing the cost of verifying their identities would reduce
the total cost of the system to a recipient.
\item In the LOCKSS case,
repeated transactions are the norm and this property
would be very useful.
\end{itemize}
\item It provides another parameter with which to adjust the
cost of the proofs requested from a generator,
namely the size of the range of indices to be searched.
In the LOCKSS case,
this helps adjust the effort requested to the economic requirements
and the underlying hash cost.
\item In effect, we have converted the uncertainty in \emph{time} inherent
to Algorithm MBound into uncertainty in \emph{size};
there is a non-zero probability that every value in the requested
range on indices has the required number of low-order zeros and
must be included in the proof.
There is no upper bound on the time required by Algorithm MBound,
there is none on the message size required by our new scheme.
\end{itemize}

Note that
in this scheme there may not be any indices $i$ in the requested
range which result in values with $e - m$ low-order zeros.
Thus an empty list of indices is potentially a valid proof of
effort,
albeit one that is anomalously expensive to verify.
If an empty proof is valid,
verifying it takes as much effort as generating it.
If it is invalid,
the first valid path found in the range shows this.
On average this takes $1/{2^m}$ of the effort that should have been
used to generate the proof.
This poses a problem;
an adversary can generate empty proofs free and cause verifiers to
waste on average $2^{(e-m)}$ trials to reveal they are invalid:

\begin{itemize}
\item In the spam case,
the recipient should reject all empty proofs.
A sender finding an empty proof should perturb the message and try again.
\item In the LOCKSS case,
the requestee should reject all empty proofs.
If the parameters can be set to make empty proofs sufficiently rare,
the fault tolerant nature of the protocol means that this will have
little effect.
\end{itemize}

\section{\label{sec:settingParameters}Setting Parameters}

The authors chose parameters for Algorithm MBound that prove $2^{15}$ paths
each $2^{12}$ steps long, and thus $2^{27}$ cache misses.
To choose comparable parameters for our scheme,
we start by setting $m = 4$ to keep the average proof size
to 16 indices or 64 bytes.
If we use the same path step mechanism as Algorithm MBound,
we also need to set $l = 2048$ and thus need a range of $2^e = 2^{15}$ indices
to search.
We thus look for paths generating a value with $(e - m) = 11$ low-order zeros.
With these choices,
the probability that all $2^{15}$ trials fail to find any such path is
$(1 - 1/{2^4})^{2^{15}}$ or about $10^{-7}$.




\section{\label{sec:futureWork}Future Work}

We have yet to implement our new scheme in the LOCKSS protocol.
We expect it to resolve many tricky issues that arose in our
attempt to use Algorithm MBound.  Doubtless,
other tricky issues will arise as we proceed.
It may well be that the path generation part of Algorithm MBound
is not the best choice for
generating the larger number of smaller sub-proofs
that our scheme needs;
this is the subject of further exploration.

\section{\label{sec:conclusion}Conclusion}

We have shown that the actual cost, measured in the number of trials,
needed to generate a proof of effort using Algorithm MBound
has a fairly broad distribution.
We have identified a number of design issues this raises for a
practical application of this scheme to both its intended spam
prevention purpose,
and for adjusting the cost of operations in the LOCKSS protocol.
We suggest an alternative scheme,
in which the cost of generating a proof is fixed and the cost of
verifying a proof can be adjusted by the verifier to match his
suspicions of the generator.

\section{\label{sec:acknowledgements}Acknowledgements}

This material is based upon work supported
by the National Science Foundation under Grant No.\ 9907296, however any
opinions, findings, and conclusions or recommendations expressed in this
material are those of the author and do not necessarily reflect the
views of the National Science Foundation. 

The LOCKSS program is grateful for support from the National Science
Foundation,  the Andrew W. Mellon Foundation, Sun Microsystems
Laboratories, and Stanford Libraries.

Petros Maniatis, Mema Roussopoulos, TJ Giuli and Mary Baker provided
many helpful comments.



\bibliographystyle{plain}
\bibliography{../common/bibliography}

\begin{thebibliography}{1}

\bibitem{Abadi2003}
Mart\'{\i}n Abadi, Mike Burrows, Mark Manasse, and Ted Wobber.
\newblock {Moderately Hard, Memory-bound Functions}.
\newblock In {\em Proceedings of the 10th Annual Network and Distributed System
  Security Symposium (NDSS)}, San Diego, {CA}, {USA}, February 2003.

\bibitem{Dwork2003}
Cynthia Dwork, Andrew Goldberg, and Moni Naor.
\newblock {On Memory-Bound Functions for Fighting Spam}.
\newblock In {\em Advances on Cryptology (CRYPTO 2003)}, Santa Barbara, {CA},
  {USA}, August 2003.
\newblock To appear.

\bibitem{Maniatis2003lockssSOSP}
Petros Maniatis, Mema Roussopoulos, TJ~Giuli, David S.~H. Rosenthal, Mary
  Baker, and Yanto Muliadi.
\newblock {Preserving Peer Replicas By Rate-Limited Sampled Voting}.
\newblock In {\em Proceedings of the Nineteenth {ACM} Symposium on Operating
  Systems Principles}, pages 44--59, Bolton Landing, {NY}, {USA}, October 2003.

\end{thebibliography}

\section*{\label{sec:Appendix}Appendix}

We used the abstract model implemented by the program below to
generate the data for the figures.
We had two reasons for this:
\begin{itemize}
\item Running large numbers of effort proofs is time-consuming.
\item We wished to demonstrate that the problems we identify are
a consequence of the way the proof search is organized and not of
the particular cryptographic features of Algorithm MBound.
\end{itemize}
The program generates graphs similar to those of an actual implemenation
of Algorithm MBound for low $e$.

{
\small
\begin{verbatim}
import java.text.*;
import java.io.*;

public class MBFcost {
  static double[] numTry = new double[32];
  static double[] costTry = new double[32];
  static double totalCost = 0;
  static long e = 0;
  static double p = 0.0;
  static int maxTries = 0;

  public static void main(String args[]) {
    long m = (long)Integer.parseInt(args[0]);
    for (e = 3; e <= m; e += 3) {
      initialize();
      fillHistogram();
      outputHistogram();
    }
  }

  static void initialize() {
    for (int i = 0; i < numTry.length; i++) {
      numTry[i] = 0.0;
      costTry[i] = 0.0;
    }
    p = 1.0/((double) (1 << e));
    maxTries = (1 << (e + 8));
  }

  static void fillHistogram() {
    double probSoFar = 1.0;
    for (int i = 1; i < maxTries; i++) {
      int b = bin(i);
      // The probability of a run ending
      // on try i is 1/2^e times probability
      // of not having ended before
      probSoFar *= (1.0 - p);
      double probHere = p * probSoFar;
      numTry[b] += probHere;
      // Thus the cost of a run ending on
      // try i is i/2^e
      double c = ((double) i) * probHere;
      costTry[b] += c;
      totalCost += c;
    }
  }

  static void outputHistogram() {
    NumberFormat nf =
      NumberFormat.getNumberInstance();
    nf.setMinimumFractionDigits(2);
    try {
      FileOutputStream numFOS =
        new FileOutputStream("data/e" + e +
                             "/tries.dat", true);
      PrintWriter numberPW = new PrintWriter(numFOS);
      FileOutputStream costFOS =
        new FileOutputStream("data/e" + e +
                             "/cost.dat", true);
      PrintWriter costPW = new PrintWriter(costFOS);
      for (int i = 0; i < (e + 8); i++) {
        String pcTry = nf.format(numTry[i] * 100);
        String pcCost =
          nf.format((100*costTry[i])/(totalCost));
        numberPW.println(i + "\t" + pcTry);
        costPW.println(i + "\t" + pcCost);
      }
      numberPW.close();
      costPW.close();
    } catch (FileNotFoundException ex) {
      // No action intended
    }
    
  }

  static int bin(int p) {
    int ret = 0;
    if (p < 0)
      p = -p;
    while (p != 1) {
      ret++;
      p >>>= 1;
    }
    return ret;
  }
    
}
\end{verbatim}
}

\end{document}